# Description of dipole strength in heavy nuclei in conformity with their quadrupole degrees of freedom

E. Grosse[1,2,a], A.R. Junghans[1,b], R. Massarczyk[1], R. Schwengner[1] and G. Schramm[1]

[1] Helmholtz Zentrum Dresden-Rossendorf (HZDR), Germany
[2] Technische Universität Dresden, Germany

**Abstract.** In conformity to new findings about the widespread occurrence of triaxiality arguments are given in favor of a description of the giant dipole resonance in heavy nuclei by the sum of three Lorentzians. This TLO parameterization allows a strict use of resonance widths Γ in accordance to the theoretically founded power law relation to the resonance energy. No additional variation of Γ with the photon energy and no violation of the sum rule are necessary to obtain a good agreement to nuclear photo-effect, photon scattering and radiative capture data. Photon strength other than E1 has a small effect, but the influence of the level density on photon emission probabilities needs further investigation.

## 1 Parameterization of the isovector giant dipole resonance (IVGDR)

The width of the isovector giant dipole resonance (IVGDR) plays an important role in predictions for the electric dipole strength in heavy nuclei. Based on the presumption that the tail of the IVGDR determines this strength down to low energies a good knowledge of this parameter is essential: Away from the maximum the height of a Lorentzian is nearly directly proportional to its width. Cross sections for compound nucleus reactions with photons are thus strongly depending on it. Attempts to obtain values for this width Γ by Lorentzian fits to the IVGDR peak region for each nucleus individually have resulted [1, 2] in surprisingly strong variations of Γ with A and Z – as e.g. presented [3, 4] within RIPL-3. This is shown in Figure 1 where it is also shown that the scatter is reduced when the splitting of the IVGDR in strongly deformed nuclei is accounted for by using the sum of two Lorentzians for the fit.

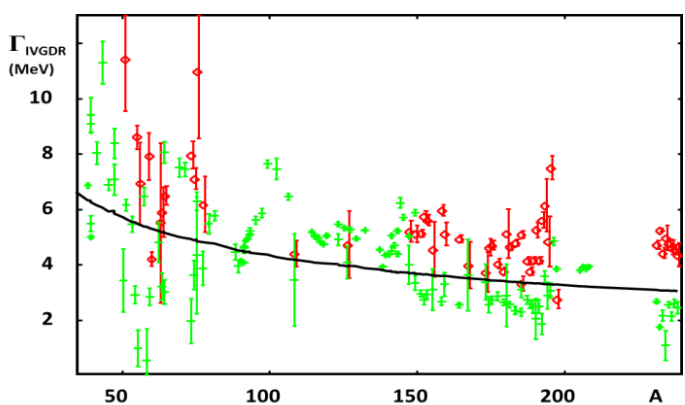

**Fig. 1.** Variation of the apparent width of the IVGDR as obtained from a fit [4] of Lorentzians to absorption data. If two Lorentzians were used, the width of the higher energy one is shown in red and green corresponds to the lower, resp. the only one. The black curve depicts the width predicted by the power law [8] between resonance energy [10] and width.

[a] e.grosse@hzdr.de, [b] a.junghans@hzdr.de

For the many nuclei neither well deformed nor spherical a wide scatter is observed [4, 5], which appears rather erratic not displaying a clear systematics. In contrast, many theoretical descriptions of collective nuclear degrees of freedom have postulated a direct relation between the width $\Gamma$ of the IVGDR and the resonance energy $E_o$, which varies smoothly with A and Z. A power law $\Gamma \propto E_o^\delta$ has been proposed [7] since long and calculations [8] based on the wall formula have determined $\delta \cong 1.6$. In Figure 1 the suggested variation of $\Gamma$ with A for $\delta = 1.6$ is also shown; this value was demonstrated [8] to be valid also for non-spherical including non-axial shapes. Similar to previous studies [6, 9] the resonance energies $E_o$ are taken from the droplet model approach [10]. Obviously the practice of extracting the IVGDR width from a fit to the photon absorption data for individual nuclei is at clear variance to the combination of the predictions for $E_o$ and $\Gamma$, both based on hydro-dynamical considerations widely accepted for the description of the IVGDR. In the following additional drawbacks of this fitting procedure will be described together with problems for the application of the dipole strength function resulting from it.

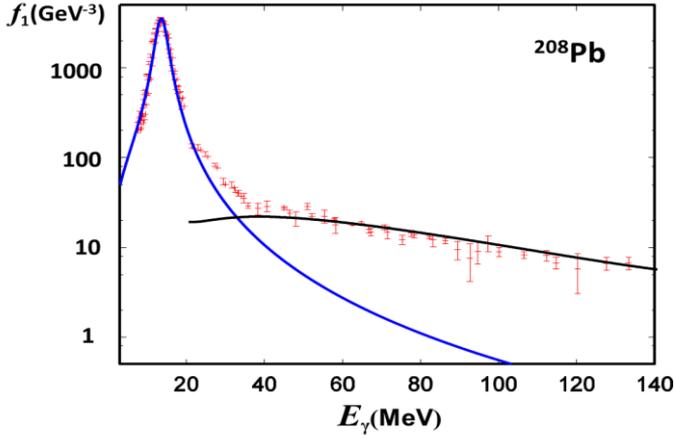

**Fig. 2.** Photon strength in $^{208}$Pb as observed [1, 15] up to 140 MeV by photo-neutron emission. The drawn lines correspond to a Lorentzian in conformity with the TRK sum rule (blue) and the prediction of the quasi-deuteron effect ([14], black).

If the integrated strength of a Lorentzian is described by $\int \sigma_{abs}\, dE = \frac{\pi}{2}\sigma_0\Gamma$ an error in $\Gamma$ causes a respective deviation in the integral of the cross section as long as the maximum is taken from the fit as well. The energy integrated electric dipole strength of nuclei is strongly dominated by the IVGDR such that a deviation there results in a corresponding change in the dipole sum. From rather general arguments [11-13] prediction for the sum rule were derived, which corresponds to the integral taken up to the energy where sub-nuclear degrees of freedom allow additional photon absorption. It can be written as:

$$\int \sigma_{abs}\, dE \cong 6.0\frac{NZ}{A} + \kappa A \quad \text{with} \quad \kappa < 0.6 \qquad (1)$$

Here the 2$^{nd}$ term is a small contribution which mainly represents the photo-dissociation above the IVGDR identified [14] as the quasi-deuteron effect. For the spherical $^{208}$Pb this is seen in Figure 2, where IVGDR data [1, 15] are depicted after correcting them for a fault in the calibration discussed [16] since long and confirmed [17, 18] recently at HZDR-ELBE.

For many non-spherical nuclei the individual fits presented [1, 2] about 30 years ago and taken up again recently [3, 4] find a considerable excess above the sum rule prediction as shown in Figure 3. This excess is apparently larger than contributions possibly related to a momentum dependence of the nucleon-nucleon interaction [12]. These should also be visible in $^{208}$Pb and other non-deformed nuclei for which, as seen in Figure 2 and 3, the absorption cross section in the IVGDR agrees to the TRK sum rule [11]. We will show this excess to disappear if a breaking of the axial symmetry is considered as well.

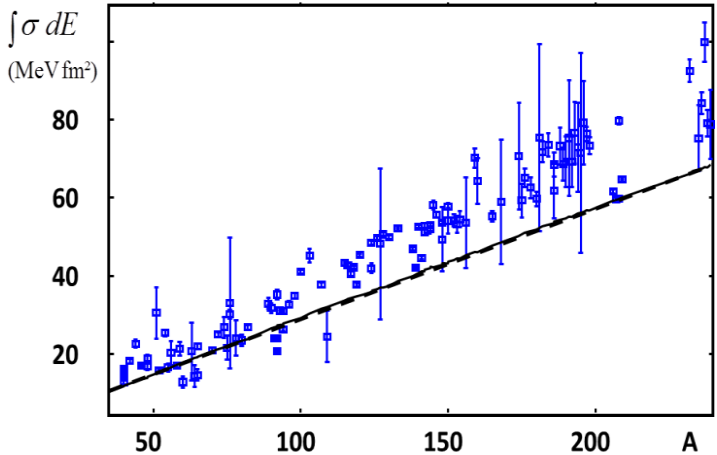

**Fig. 3.** Absorption cross sections as obtained [4] from the fit if up to two Lorentzians to the data for individual nuclei (blue squares with error bars). A clear excess above the TRK sum rule ([11], black curve) is observed for most nuclei away from magic nucleon numbers.

The low energy tail of the IVGDR has been considered [9, 19-22] since long to approximate the actual electric dipole absorption in rather good way. As the strength function in the tail region strongly influences the radiative capture of nucleons many attempts have been undertaken to derive experimental information on it. Besides neutron capture through well understood resonances and average neutron capture (ARC) resonant photon scattering can be used to derive the photon strength below the neutron separation energy $S_n$. Respective experiments have been performed with bremsstrahlung produced at various electron accelerators and results obtained [18, 23] at HZDR-ELBE are shown in Figure 4. They are depicted together with photo-neutron data [24] for the resonance region and respective results from calculations.

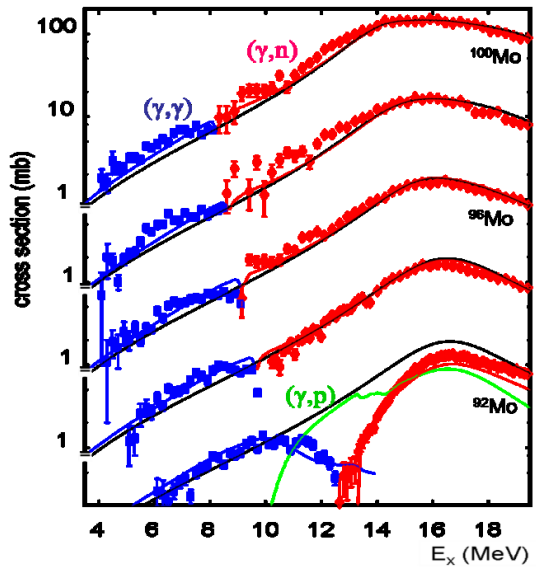

**Fig. 4.** Experimental cross sections for photon-induced processes in $^{92,94,96,98,100}$Mo (from bottom). The data at low $E_x$ from photon scattering (■, blue) [18] are shown together with (γ,n) data (♦, red) [24] rescaled as described in the text. The thin lines depict the results of Hauser-Feshbach calculations performed with the code TALYS [25] [blue: (γ,γ); red: (γ,n); green: (γ,p); only shown as long as their contribution exceeds 10%]. These calculations use the TLO-parameterization [9] for the absorption cross section $\sigma_\gamma$ (E1), represented by the thick solid line. A cross section overshooting this line indicates contributions from modes other than E1 as will be discussed in chapter 4.

Figure 4 depicts a typical example for the good match between the photon absorption derived from nuclear photo-effect data [24] and photon scattering [18], respectively. The cross sections are compared to a parameterization by a triple Lorentzian (TLO) to be described in the following. The good agreement between data and parameterization indicates that there is no need to introduce a dependence of the resonance width Γ on the photon energy or a departure of the resonance strength from the TRK sum rule. Obviously the most neutron deficient isotope emits protons already below the energy needed for the (γ,n)-reaction. As respective data are not available, this contribution is inserted as calculated by the code TALYS [25]. This code had to be modified to the TLO parameterization [9]: it was also used to include the contribution due to M1 strength from the isoscalar and isovector spin flip modes [26]. The photo-absorption cross section was obtained from the (γ,γ)-scattering yield by correcting for branching and feeding effects by a procedure similar to the one proposed [27] for two-step cascades following n-capture.

The TLO description of the IVGDR is obtained from inserting independently determined deformation parameters β and γ into the Hill-Wheeler expression [28] for the three axes of the triaxial ellipsoid:

$$E_k = E_0 \exp\left\{-\sqrt{5/4\pi}\,\beta \cos(\gamma - \tfrac{2}{3}k\pi)\right\};\quad \Gamma_k = c \cdot E_k^{1.6};\quad E_0 = E_0(J, Q, m_{eff}) \qquad (2)$$

Here it is assumed that the centroid energies of the IVGDR three dipole oscillation frequencies are inversely proportional to the axis lengths and the volume $R_1 R_2 R_3 = 1.16\,A^{1/3}$ is conserved. The widths of the three resonances are derived from the power law mentioned above and their integral strength is taken to be equal with the sum of the three given by the TRK sum rule [11] (cf. Eq. 1). The cross section is then given by the sum of the respective three Lorentzians:

$$\sigma_\gamma = \sum_{k=1,2,3} \frac{2 I_j}{\pi} \frac{E^2 \Gamma_k}{\left(E_k^2 - E^2\right)^2 + E^2 \Gamma_k^2};\quad I_1 = I_2 = I_3 \qquad (3)$$

The centroid energies $E_o$ are taken from the droplet model approach with the symmetry energy $J = 32.7$ MeV and the surface stiffness $Q = 29.2$ MeV obtained [29] from a fit to the ground state masses. One free parameter introduced here is an effective mass $m_{eff} \cong 874$ MeV and the only other one is the proportionality factor $c \cong 1/20$ MeV$^{-0.6}$, both were obtained from an adjustment to IVGDR data of more than 50 nuclei with $70 < A < 240$. The only additional information needed to arrive at a good fit – in accordance to the TRK sum rule [11] – concerns the ground state shape parameters.

## 2 Deformation and triaxiality of nuclear shapes

As seen from the different apparent widths of the IVGDR-distributions in the five isotopes of Mo, their different ground state deformation has a considerable influence. Several methods of extracting a departure from axial symmetry have identified many nuclei with $A > 70$ as triaxial and have thus confirmed earlier assumptions about non-zero triaxiality γ in nuclei, mainly in those with small quadrupolar deformation. Some time ago a method was developed to analyze multiple Coulomb excitation data via model-independent invariants [30, 31] and this has also given a clear proof of the breaking of axial symmetry; a small γ is observed [32] in nuclei showing a large intrinsic quadrupole moment $Q_o$, i.e. large deformation β.

Detailed theoretical investigations explicitly studying low energy quadrupole excitations in heavy nuclei require a considerable increase of the computational effort to be extended to full triaxiality such that not many results are available. Very recently a systematic study [33] of low energy nuclear structure was carried out at CEA and IPN Orsay together with INT at Seattle using the Gogny interaction within the Hartree-Fock-Bogoliubov (HFB) formalism allowing also triaxial shapes. Both, experiment as well as theory, show that ground states of 'transitional' nuclei are better parameterized by adding a triaxiality γ to $Q_o$, characterizing the deformation β. Experimental information on β has been reviewed [34] and it was shown, that even if the deformation parameters as used in a study do not always respect volume conservation [28] explicitly they may be plotted together without large errors. In Figure 5 a compilation of various experimental data [30, 31, 35-41] is shown (on the left) together with (on the right) results of the calculations [33] mentioned – both for nuclei near the valley of stability. The right plot also depicts the calculated effect of zero point oscillations which we have considered, as will be pointed out below.

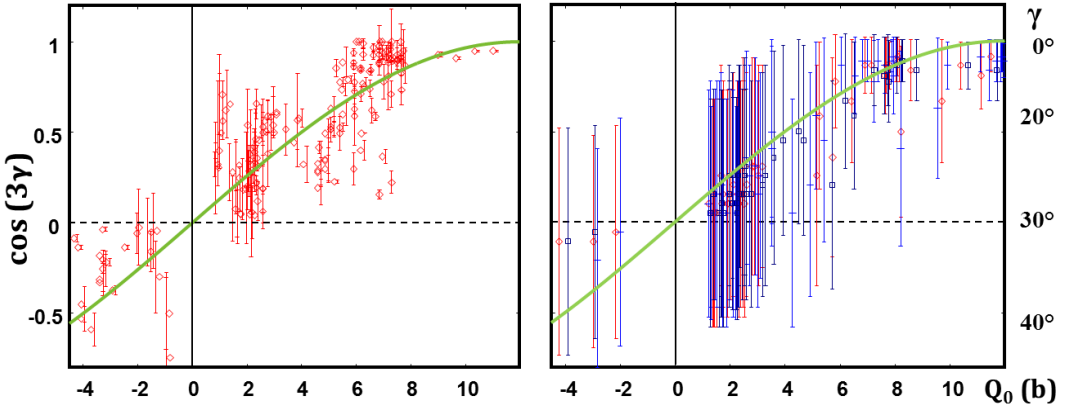

**Fig. 5.** Correlation of two parameters describing the departure from sphericity ($Q_o$) and axiality ($\cos 3\gamma$) for various nuclei in the valley of stability. The data [30, 31, 35-41] in the plot on the left are obtained by various experimental methods and the right plot depicts results of a HFB-calculation [33] also taken from the literature. Whereas the vertical bars in the left plot indicate experimental uncertainties for the determination of the triaxiality, the bars in the right plot indicate the large standard deviations due to zero point oscillations. The green curve is an eye guide to indicate the average correlation between the two deformation parameters already noted previously [41].

The rather good correlation between $\beta$ and $\gamma$ as observed [41] earlier is depicted by the green curve which may serve as an approximation for very small $\beta$ and in case only $\beta$ is known. We have used available triaxiality information to describe the IVGDR shapes by a sum of three Lorentzians (TLO) [9]. We get good agreement to photonuclear data [42] with $\Gamma \propto E_k^{1.6}$ not only in the peak but also in the tails, as obtained [43] by ARC. In Figure 6 the IVGDR is plotted for two nuclei of different deformation as dipole strength function [21]:

$$f_1(E_\gamma) = \frac{\overline{\sigma_\gamma}}{3(\pi\hbar c)^2 \overline{E_\gamma}} \qquad (4).$$

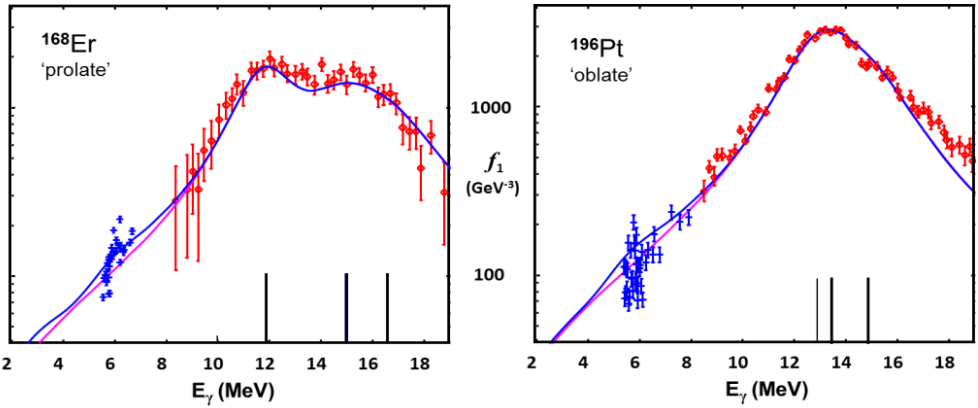

**Fig. 6.** TLO descriptions of the dipole strength in two heavy nuclei, one having a prolate shape in its ground state, the other one is oblate. As the black bars indicate, a non-axiality as indicated by the green curve in Figure 5 improves the agreement to the data [42, 43], which are taken from ($\gamma$,n) and (n,$\gamma$)-experiments (red and blue, resp.). The curve drawn in magenta depicts the E1-strength and the blue curve includes the effect of M1, as described later.

In the plot for $^{168}$Er the effect of the increase of Γ with $E_k^{1.6}$ is clearly seen: The higher energy peak is smaller in height as compared to the lower one, although it represents two dipole frequencies. Apparently Porter Thomas fluctuations [44] play a strong role in the case of small level density.

The IVGDR shapes in the even isotopes of Nd have often been shown [45, 46] as an example for the effect of the departure from sphericity. It is thus indicated to apply the TLO parameterization to them and this is done in Figure 7. Besides the deformations β and γ obtained from other sources the only two parameters applied are the ones mentioned above – and they are taken from the global fit to many nuclei with 70 < A < 240. As obvious, the agreement is compatible to the one achieved by fitting each isotope with individual (local) resonance parameters [45, 46].

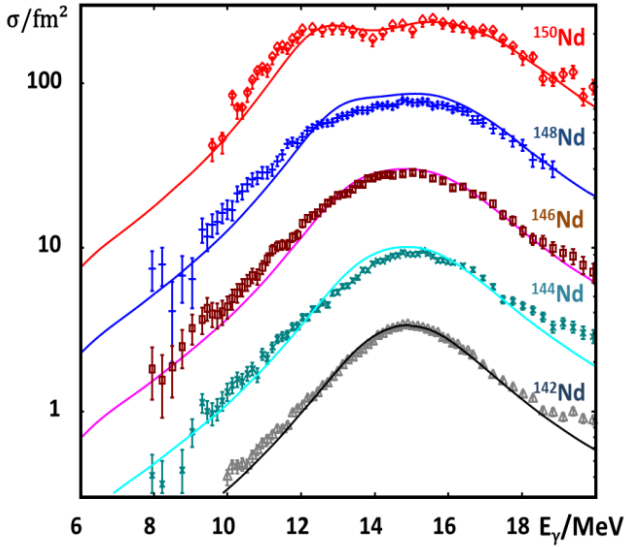

**Fig. 7.** Photo-neutron data [45] for $^{142-150}$Nd depicted in comparison to the respective prediction of the TLO parameterization. The vertical scale corresponds to $^{146}$Nd; for the other isotopes powers of √10 have to be applied.

The good agreement as demonstrated here -- as well as in a large number of other heavy nuclei with A>70 [9, 18, 45] -- allows the following conclusions:
  a) The regard of triaxiality leads to the use of a resonance width Γ in accord to a global relation to the respective resonance energy by a power law.
  b) In most cases this width is smaller as compared to the one derived by a local fit with one or two Lorentzians resulting in a reduction of the cross section in the tail region.
  c) This has a strong effect on the extrapolation to the full energy range of relevance for the dipole sum rule; no strong excess in strength above the TRK sum is observed.
  d) The agreement to average resonance capture data is satisfactory also in the low energy tail without the need for an additional dependence of Γ on the photon energy.

The often rather large experimental uncertainties in γ have no significant effect on these statements as they only influence the IVGDR shape near its maximum [18]. Earlier work in this field [49-51] based on the neglect of triaxiality and applied recently [3, 4] disagrees considerably to the conclusions made here.

To account for the large zero point oscillations especially strong for the triaxiality described by cos(3γ) we combined the TLO ansatz to the concept of instantaneous shape sampling [52]. The ground state β and γ deformations were sampled 20 times around their mean values using the predicted variances – both taken from the calculations depicted in the right part of Figure 5 – and formed averages for the calculated cross sections. The left part of Figure 8 shows the result for $^{106}$Pd in comparison to data [48] for Pd(γ,n). The small influence on radiative neutron capture becomes obvious when regarding a schematic prediction for the spectrum of the 1$^{st}$ gamma ray after capture -- also shown in this panel.

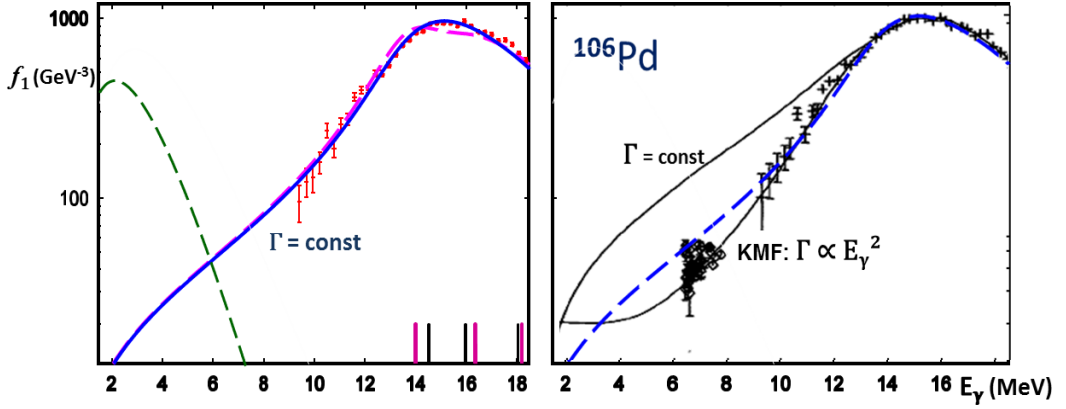

**Fig. 8 left:** Data for Pb(γ,n) in comparison to the TLO prediction for $^{106}$Pd without (dashed purple) and with shape sampling (blue curve) together with a sensitivity curve for radiative neutron capture (dashed green).
**Right:** The same data in a plot comparing a previously made fit assuming sphericity and a change of the width Γ with $E_\gamma^2$ (lower curve) to a fit corresponding to Γ not depending on the photon energy (upper curve) [49].

The right part of the Figure is taken from a single Lorentzian fit to the IVGDR peak area as published previously [49], which apparently overestimates the width Γ. This had suggested the inclusion of an explicit photon energy dependence [49-51] which, if perpetuated to higher energy, would enhance the integral by up to 80 % above the TRK sum. The introduction of triaxiality as proposed here (TLO) achieves good agreement with smaller Γ in agreement to the power law. As this has an appreciable influence on the energy range with high sensitivity to gamma decay (cf. left panel of Figure 8) it is thus of interest to regard compound nucleus reactions with photons in the exit channel.

## 3 Photon strength and radiative capture

A large number of data on photon emission exists for the radiative capture of neutrons in the keV range. Time of flight experiments have produced results for total and for radiative widths of various resonances [3]. A quantity generally used to describe their gamma decay is the average radiative photon width <Γ$_\gamma$> of a resonance $R$ at energy $E_R$. It can be calculated from the photon strength function $f_1(E_\gamma)$:

$$\langle \Gamma_\gamma \rangle = \left\langle \sum_f \Gamma_\gamma(R \to f) \right\rangle = 3 \int_0^{E_R} \frac{\rho(E_f)}{\rho(E_R)} E_\gamma^3 f_1(E_\gamma) dE_\gamma \qquad (5).$$

Here the Axel-Brink hypothesis [19-21] is used, which relates the photon strength function derived from absorption data (Eq. 4) to gamma decay widths:

$$f_1 E_\gamma^3 = \langle \Gamma_\gamma \rangle \rho(E_R) \qquad (6).$$

The ratio of the spin distributions of the final and initial level densities is approximated by 3. As obvious from Eq. 5 the level density $\rho$ enters only as a ratio for the resonance energy $E_R$ and the final energy $E_f = E_R - E_\gamma$. If the level densities are taken from the constant temperature model (CTM) this ratio is already fixed by the parameter T, which we approximated by (using a given A-dependence [53]):

$$T \cong 17.6 \text{ MeV} \cdot A^{-2/3} \qquad (7).$$

With these approximations the proposed TLO parameterization can be tested globally by a comparison to average radiative widths obtained [3] from the analysis of neutron resonances in many nuclei. This is shown for s-wave resonances in Figure 9:

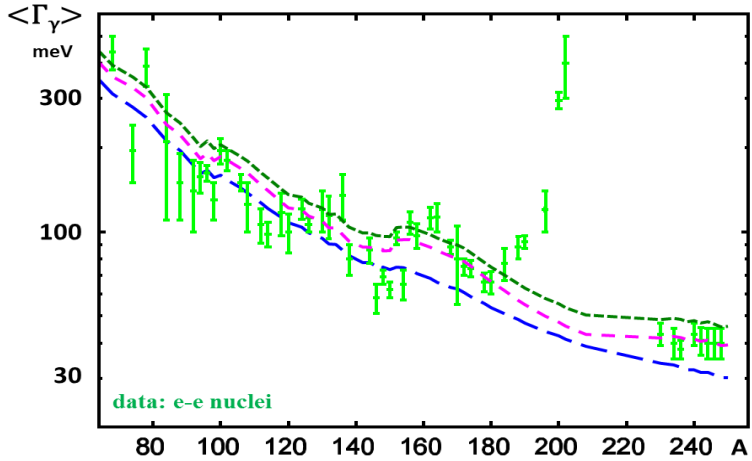

**Fig. 9.** Experimentally determined [3, 54] average radiative widths for even nuclei with 70<A<240 (green bars) in comparison to predictions obtained from TLO (blue long dashed curve). The short dashed green curve was obtained by adding several other contributions to the E1 prediction of TLO (see text). The medium dashed red curve results, if only M1 strength originating from spin flip and orbital magnetic modes [26] is included.

The Figure demonstrates that the large variations in electric dipole strength and level density between A~80 and A~240 are partly compensating each other such that <$\Gamma_\gamma$> varies only little in that mass range – as is well known from observations and clearly documented by the data compilation [3, 54]. Obviously the global parameters used here work for the majority of the heavy nuclei, but additional effects have to be considered for near magic nuclei with their small level density. In the Figure an approximate account is made for various contributions to the photon strength which have been discussed in the literature recently:

a) In deformed nuclei a scissors like vibrational mode [26] induces enhanced M1 transitions at energies well within the range of high sensitivity as shown in Figure 8 left.
b) Magnetic strength originating from isoscalar and isovector spin-flip modes [26] is occurring in an higher energy range, as reviewed recently (together with the scissors mode).
c) Experimental data [55] indicate the presence of dipole strength even at such low energy where a Lorentzian falls to zero; M1 strength is possible for $E_\gamma \rightarrow 0$, as are nuclear magnetic moments.
d) Studies [56, 57] combining information from photon and α-particle scattering claim to have identified isoscalar electric strength at energies near half the IVGDR energy.
e) Lower energy E1 strength well in the sensitivity window has been observed [58, 59] in many nuclei and was assigned to the coupling of quadrupole to octupole vibrations.

In principle all these have to be taken into account to arrive at a proper description of radiative processes. Estimates indicate that each of the five modes contributes at most a few percent to the total sum, but the effect at low energy may be higher. From the two dashed curves in Figure 9 it can be concluded that even the sum of all will contribute less than the tail of the isovector electric dipole mode. Dedicated studies are necessary to further specify this finding and it has to be clarified, if for all the five extra contributions the Axel-Brink hypothesis can be applied; this point is important for the use of the respective strength functions for multiple cascade decays [27].

For applications in nuclear astrophysics and nuclear power technology a reliable prediction of cross sections for the radiative capture of neutrons in the keV to MeV energy range is of great interest. It is thus interesting to expand the global parameterization presented so far to this problem. In the following first results for comparisons of TLO-predictions to published capture data will be presented. A quasi-classical expression [60, 61] for $\sigma(n,\gamma)$ is:

$$\langle\sigma(n,\gamma)\rangle \approx 2\pi^2(\lambdabar_n+r)^2 \cdot \rho(E_R)\langle\Gamma_\gamma\rangle \qquad (8)$$

where $\lambdabar_n$ stands for the neutron wavelength and $r$ is the radius of the target nucleus. The level density at the capture resonance at $E_R = S_n+E_n$ hardly differs from the one at $S_n$ and can be approximated by

$$D(S_n) = 1/\rho \cong 8 \text{ MeV} \cdot e^{-S_n/T} \qquad (9)$$

if no tabulated vales are available – which is the usual case for a global parameterization.

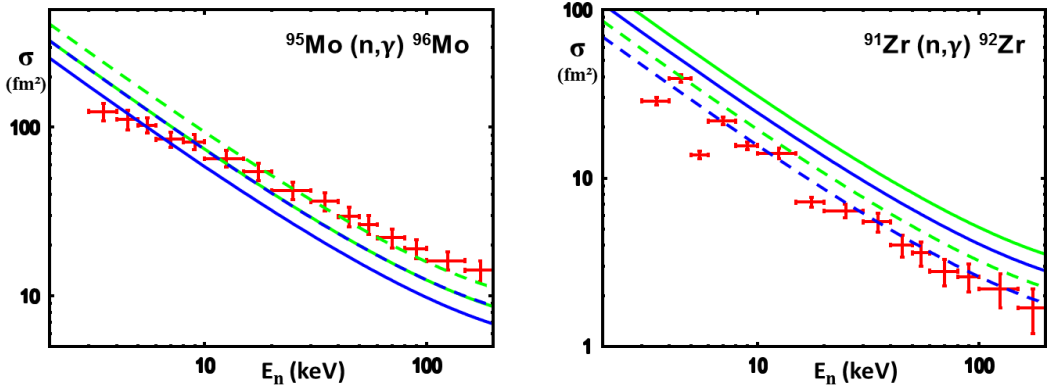

**Fig. 10.** Experimental (n,γ) cross sections [62] in comparison to Eq. 8 as based on Eq. 5 and TLO (blue curve) and with inclusion of M1 modes a-c. The drawn lines are calculated with global T and ρ, the dashed ones with tabulated values from RIPL-3 [3].

Figure 10 shows radiative neutron capture cross sections calculated using the described global parameterization in comparison to experiments [62] for nuclei near the N=50 neutron shell closure. Nuclei in this region of the nuclide chart had been studied in detail at HZDR-ELBE (cf. Figure 4) and the TLO parameterization was originally developed on the basis of these studies. The advantage of it for nuclei which are neither spherical nor well deformed was worked out in connection to Figures 6-8; it is hence of interest to test it as well for such nuclides. In Figure 11 the case of $^{156}$Gd is presented, a typical case of a nucleus generally assumed to be axially deformed.

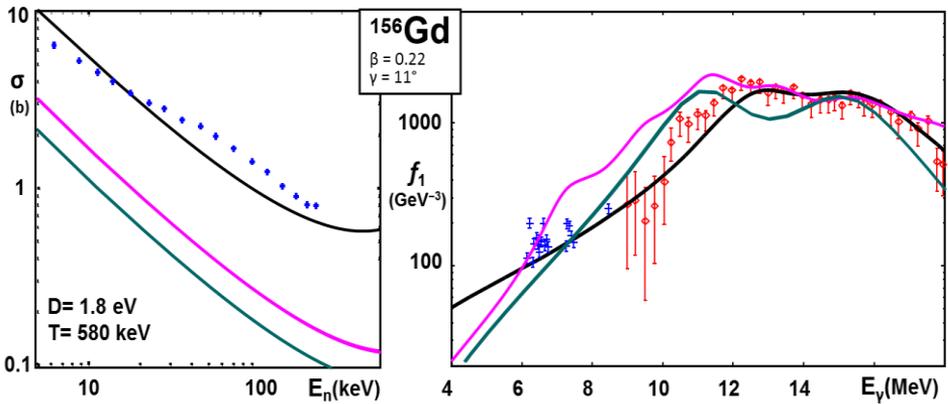

**Fig. 11.** $^{156}$Gd as studied by radiative capture [43, 63] (blue +) and by photon absorption [64] (red ⊗). Shown for comparison are predictions made on the basis of the CTM for level densities combined to the EGLO parameterization [3] for the photon strength (lower cyan curve) as well as to a QRPA-calculation [3] (middle magenta curve). The best agreement to the experimental data is seen for the top black curve as derived from the TLO photon strength function.

A comparison of radiative capture and photon absorption data to predictions based on TLO as well as on QRPA-calculations [3] with a density dependent NN-interaction (SLy4) and on a generalized double Lorentzian (EGLO)[3] clearly indicates deficiencies for the two latter schemes in comparison to TLO. At variance to the original proposal [3] EGLO is used in Figures 11 and 12 such that it accords to the Axel-Brink hypothesis [19-21] – no difference between absorption and decay is allowed for.

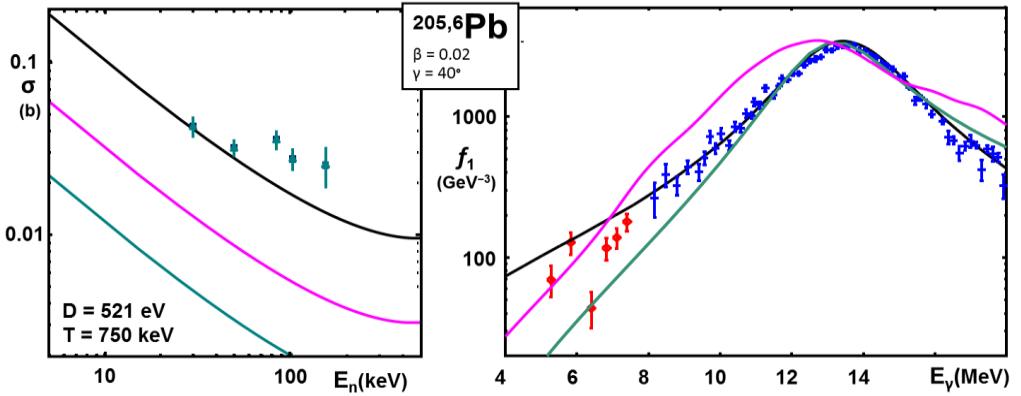

**Fig. 12.** Data [65] for $^{204}$Pb(n,γ) are depicted in the left panel in comparison to predictions made on the basis of the CTM for level densities. These are combined to the EGLO parameterization [3] for the photon strength (lower cyan curve) as well as to a QRPA-calculation [3] (middle magenta curve). The top black curve as derived from the TLO shows the best agreement to the experimental data. This is also true for the comparison to data for $^{206}$Pb(γ,n) [66] (blue +) and for $^{206}$Pb(γ,γ) [67] (red ⌥) presented in the right panel (the curve for TLO is the top one at 4 MeV).

Similar conclusions can be drawn when looking at Pb-isotopes, the prime examples for sphericity The data for $^{204}$Pb(n,γ) and for $^{206}$Pb(γ-abs) are shown in Figure 12; the latter can be considered to well represent $^{205}$Pb, which is unstable and not studied yet with photons. Again the comparison to the three predictions clearly favors the TLO parameterization, when combined to the CTM for the average level density ρ. The calculated cross sections for radiative neutron capture are directly proportional to ρ such that independent level density information for the energy region at and below the binding energy $S_n$ is essential for a full quantitative understanding of the photon emission by compound nuclei.

## 4 Conclusions

A description of electric dipole strength in heavy nuclei was presented which is in conformity with their quadrupole degrees of freedom. New spectroscopic information from Coulomb excitation and other experimental studies was shown to agree to a recent theoretical investigation using the Hartree-Fock-Bogoliubov scheme to calculate low energy properties of heavy nuclei: A hitherto often neglected fact is the symmetry breaking at and near the ground states in most heavy nuclei. One observes not only the loss of spherical but also of axial symmetry and hence the presence of a triaxial nuclear shapes. The existence of three different body axes leads to three different frequencies for the giant dipole vibrational mode, an effect which is difficult to observe directly in the photon absorption spectra because of the large spreading width Γ of the giant dipole resonance. But when the information about the nuclear shape as available from other sources is accounted for a good description of the form of the IVGDR is possible. It often has the additional advantage, that the widths of the three components are smaller than the apparent breadth with the consequence of the disappearance of a significant departure from the TRK sum rule. A second effect is the now acceptable strict proportionality of Γ to a theoretically predefined power of the resonance energy, and the proportionality constant is the same for the three parts in one nucleus as well as for the many nuclei investigated in the range 70 < A < 240.

The good description of photon absorption data by the triple Lorentzian (TLO) parameterization inspires to apply it to radiative capture processes as well. A first test here is performed by the comparison to experimental average radiative widths determined from neutron resonances and it shows good agreement for many non-spherical nuclei if the level densities are taken from the constant temperature model CTM. The also observable good match between radiative capture data and the TLO prediction suggests the use of it as a basis for a scheme to make globally valid predictions for cases where capture data are not yet available. Such a situation is given for the r- and s-processes in the cosmic element production and other scenarios of nuclear astrophysics. Radiative capture reactions in nuclear fuel, nuclear waste and in related construction materials are of importance for nuclear technology applied to waste transmutation and to alternative reactor concepts.

## Acknowledgements

Part of our research is connected to respective EURATOM funded programs such as the FP7 project ERINDA (www.erinda.org). The respective support of the work presented here is gratefully acknowledged.